# Synthesized complex-frequency excitation for ultrasensitive molecular sensing

Kebo Zeng[1]†, Chenchen Wu[2]†, Xiangdong Guo[1,2]†, Fuxin Guan [1]†, Yu Duan[2], Lauren L Zhang[3], Xiaoxia Yang[2]*, Na Liu[4]*, Qing Dai[2]* and Shuang Zhang[1,5]*

[1]New Cornerstone Science Laboratory, Department of Physics, University of Hong Kong, Hong Kong, China

[2]CAS Key Laboratory of Nanophotonic Materials and Devices, CAS Key Laboratory of Standardization and Measurement for Nanotechnology, CAS Center for Excellence in Nanoscience, National Center for Nanoscience and Technology, Beijing, China

[3]Department of Chemistry & Chemical Biology, Harvard University, Cambridge, MA 02138, USA

[4]2nd Physics Institute, University of Stuttgart, Pfaffenwaldring 57, 70569, Stuttgart, Germany

[5]Department of Electrical & Electronic Engineering, University of Hong Kong, Hong Kong, China

† These authors contributed equally to this work.

*Corresponding author: yangxx@nanoctr.cn, na.liu@pi2.uni-stuttgart.de, daiq@nanoctr.cn, shuzhang@hku.hk

**Abstract:**

Detecting trace molecules remains a significant challenge. Surface-enhanced infrared absorption (SEIRA) based on plasmonic nanostructures, particularly graphene, has emerged as a promising approach to enhance sensing sensitivity. While graphene-based SEIRA offers advantages such as ultrahigh sensitivity and active tunability, intrinsic molecular damping weakens the interaction between vibrational modes and plasmons. Here, we demonstrate ultrahigh-sensitive molecular sensing based on synthesized complex-frequency waves (CFW). Our experiment shows that CFW can amplify the molecular signals (~1.2-nm-thick silk protein layer) detected by graphene-based sensor by at least an order of magnitude and can be universally applied to molecular sensing in different phases. Our approach is highly scalable and can facilitate the investigation of light-matter interactions, enabling diverse potential applications in fields such as optical spectroscopy, metasurfaces, optoelectronics, biomedicine and pharmaceutics.

**Introduction**

Sensors have emerged as indispensable analytical tools across a wide range of important fields, encompassing environmental monitoring, food safety, and public health[1-5]. They facilitate early disease diagnosis, personalized medicine, and rapid detection of toxic agents[6-9]. However, significant challenges still exist in the effective detection of trace molecules, hindering the further development of sensors in these applications[6,10,11]. Many efforts have been made to improve the sensor sensitivity. Among the various methods explored, optical biosensors based on surface-enhanced infrared absorption (SEIRA) have attracted much attention due to their label-free nature, molecular specificity, and noninvasive performance[9,12-14]. Through strong light-matter interactions achieved by surface-plasmon polaritons (SPPs), SEIRA can enhance the detection sensitivity of the molecular vibrational fingerprints in the infrared (IR) region[15-17]. SEIRA was first demonstrated in 1980 using Ag and Au thin films[18]. However, it was not widely adopted due to the limitation of nanofabrication techniques at the time[19]. The advancement of nanofabrication and new plasmonic materials (e.g., graphene, Ge, Si, oxides, and carbon nanotubes (CNTs)) have led to the revitalization of the research in this area in recent years[13]. In particular, plasmonic nanostructures have been proved to possess much greater enhancement of biomolecule signals than metallic thin films[17].

Compared to metal-based SEIRA, strong field confinement supported by two-dimensional (2D) Dirac fermion electronic states enables graphene-based SEIRA with excellent performance in molecular characterization for gas[20] and solid phase sensing[1,21,22]. Graphene can also enhance molecular IR absorption in aqueous solution[23]. Most importantly, active tunability of graphene plasmons broadens their detection frequency range for different molecular vibrational modes by changing the doping level via gate voltage[16,24,25]. These advantages make graphene-based SEIRA a unique platform for single-molecule detection. However, the intrinsic molecular damping largely reduces the interaction between the vibrational modes and plasmons. As a result, at lower concentrations, the spectra of plasmon-enhanced molecular signals become weaker and broader, and ultimately are overshadowed by noise.

One way to compensate for molecular damping is to add optical gain materials[26-28]. However, this requires a complex setup which may not be compatible with the detection system. In addition, gain materials usually increase instability and noise[29,30]. Another way is to use complex-frequency waves (CFW); theoretical studies have proved that CFW with temporal attenuation can restore information loss due to material losses[31,32]. However, producing CFW in real optical systems remains a challenging task. A novel method for synthesizing CFW has recently been proposed[33]. This method involves treating CFW as a coherent combination of multiple real-frequency waves based on the concept of the Fourier transform. This multi-frequency approach has been recently applied to superimaging[33] and shown remarkable improvement in the imaging resolution, but its application to sensors has not yet been attempted.

Here, we demonstrate dramatic signal enhancements of the molecular vibrational fingerprints governed by synthesized CFW. We first theoretically confirm that the truncated CFW synthesized by discretized real frequency waves in a limited range can effectively compensate for molecular damping, significantly improving trace molecular signals (~1.2-nm-thick silk protein layer). Synthesized CFWs are successfully applied to enhancing the molecular signals in the mid-IR extinction spectrum for biomolecules under different conditions, including direct measurement of multiple vibrational modes of deoxynivalenol (DON) molecules and graphene-based SEIRA of proteins in the solid phase and aqueous solution. The results show that our method can improve the sensitivity of various sensors by almost an order of magnitude and advance the quantitative detection of molecules.

**Theoretical mechanism**

Without loss of generality, we model a molecular layer using the Drude-Lorentz dispersion,

$$\varepsilon(\omega) = 1 + \sum_m \frac{\omega_{\mathrm{p}m}^2}{\omega_m^2 - \omega^2 - \mathrm{i}\gamma_m\omega} \tag{1}$$

For simplicity, we assume the molecular layer has two vibrational modes, where the plasma frequencies $\omega_{\mathrm{p1}} = \omega_{\mathrm{p2}} = 128 \text{ cm}^{-1}$, the damping rates $\gamma_1 = \gamma_2 = \gamma_{\mathrm{M}} = 60 \text{ cm}^{-1}$, and the resonant frequencies $\omega_1 = 1553 \text{ cm}^{-1}$, $\omega_2 = 1666 \text{ cm}^{-1}$. Using finite-element method (FEM) simulation, we obtain the extinction spectra of the molecular layer (the light blue curve in Fig. 1a). Obviously, the key to making the resonant peaks more pronounced is to reduce the damping rate $\gamma_{\mathrm{M}}$. If $\omega$ is replaced by a complex frequency $\widetilde{\omega} = \omega - \mathrm{i}\gamma_{\mathrm{M}}/2$, the permittivity of the molecular layer becomes a purely real value $\varepsilon(\widetilde{\omega}) = 1 + \frac{\omega_{\mathrm{p1}}^2}{(\omega_1^2 - \omega^2 - \gamma_{\mathrm{M}}^2/4)} + \frac{\omega_{\mathrm{p1}}^2}{(\omega_2^2 - \omega^2 - \gamma_{\mathrm{M}}^2/4)}$. This shows that CFW with suitable temporal attenuation can fully compensate for the damping of molecular vibrational modes. Due to the difficulty of generating CFW directly, we use a new method to synthesize the truncated CFW expressed as $E_T(t_0) = E_0 \mathrm{e}^{-i\widetilde{\omega}t_0}\theta(t_0)$, where $\widetilde{\omega} = \omega - \mathrm{i}\tau/2$, and $\tau > 0$ represents temporal attenuation. $\theta(t)$ is the time truncation function to avoid energy divergence, where $\theta(t_0) = 1$ for $t_0 \geq 0$, and $\theta(t_0) = 0$ for $t_0 < 0$. Note that the time truncation will lead to appearance of sidebands around the resonances in the complex frequency spectra, which can be eliminated via appropriate average of the signal over time (Supplementary Information Note II). Based on the Fourier transform, $E_T(t_0)$ can be expanded into the integral of the real frequency components: $E_T(t_0) = \frac{E_0}{2\pi}\int_{-\infty}^{+\infty}\frac{1}{\mathrm{i}(\widetilde{\omega}-\omega')}\mathrm{e}^{-i\omega't_0}\mathrm{d}\omega'$, where $1/\mathrm{i}(\widetilde{\omega}-\omega')$ is the Fourier coefficient. Naturally, any response in the system excited by the truncated CFW can be expressed as the integral of the real frequency response $F(\widetilde{\omega}) \approx \int_{-\infty}^{+\infty} F(\omega')\frac{1}{\mathrm{i}(\widetilde{\omega}-\omega')}\mathrm{e}^{\mathrm{i}(\widetilde{\omega}-\omega')t_0}\mathrm{d}\omega'/2\pi$ in the quasi-steady state. In reality, for a sufficiently wide spectrum range, the integral can be discretized as,

$$F(\widetilde{\omega}) \approx \sum_n F(\omega_n)\frac{1}{\mathrm{i}(\widetilde{\omega}-\omega_n)}\mathrm{e}^{\mathrm{i}(\widetilde{\omega}-\omega_n)t_0}\Delta\omega/2\pi \tag{2}$$

Subsequently, we use equation (2) to calculate the extinction of the molecular layer at CFW. The extinction is represented as $I(\omega) = 1 - |t_{M}|^2$, where $t_{M} = t/t_s$, and $t$, $t_s$ are the transmission coefficients through the substrate with and without the molecular layer, respectively. For thin layer systems, $t_{M}$ can be approximated as[34],

$$t_{M}(\omega) \approx \frac{1}{1 - iP(\omega)} \tag{3}$$

Where $P(\omega) = \frac{\chi_e(\omega)\omega d}{(n_s+1)c}$, $n_s$ is the refractive index of the substrate, $d$ is the molecular layer thickness and $\chi_e(\omega)$ is the effective susceptibility. Considering the difficulty of phase measurement in practice, we can extract the phase $\arg(t_{M})$ from the amplitude $|t_{M}|$ through Kramers–Kronig relations[35] (see details in Figure S1),

$$\arg(t_{M}(\omega)) = -\frac{1}{\pi}\mathcal{P}\int_{\mathbb{R}}\frac{\ln|t_{M}(\omega)|}{\omega - \omega'}\,d\omega' \tag{4}$$

and then $P(\omega)$ can be deduced from $t_{M}$ using equation (3).

$$P(\omega) = i(\frac{1}{t_{M}(\omega)} - 1) \tag{5}$$

Note that, similarly to equation (2), equation (4) can be discretized in actual calculation. Hence, the extinction $I(\widetilde{\omega})$ for a CFW can be obtained by calculating the response $P(\widetilde{\omega})$ from $P(\omega)$,

$$P(\widetilde{\omega}) \approx \sum_n P(\omega_n)\frac{1}{i(\widetilde{\omega} - \omega_n)}e^{i(\widetilde{\omega} - \omega_n)t_0}\Delta\omega/2\pi \tag{6}$$

$$I(\widetilde{\omega}) = 1 - \frac{1}{|1 - iP(\widetilde{\omega})|^2} \tag{7}$$

Here we do not directly calculate $t_{M}(\widetilde{\omega})$ to obtain $I(\widetilde{\omega})$ because $|t_{M}(\omega)| \to 1$ as $\omega \to \infty$, which would cause relatively large errors in equation (2) from outside the finite frequency range. On the contrary, $|P(\omega)| \to 0$ as $\omega \to \infty$, making the numerical errors of $P(\widetilde{\omega})$ smaller. Further, we time-average $P(\widetilde{\omega})$ to reduce the error caused by the truncation of the CFW, the limited frequency range and discretization of frequencies (see details in Supplementary Material Note II). Accordingly, we set $\widetilde{\omega} = \omega - i\gamma_{M}/2$ to demonstrate the enhancement effect of CFW. Compared to the original signal $I(\omega)$, the resonant peaks of $I(\widetilde{\omega})$ (the dark blue curve in Fig. 1a) are significantly narrowed, which means that synthesized CFW can directly enhance the molecular vibrational

fingerprints without additional assistance.

At very low concentrations, the absolute response of the molecular layer would be too small to measure, so SEIRA is used to solve this issue. Here, we consider a graphene nanoribbon array with a period of $\Lambda = 200$ nm, and ribbon width $w = 80$ nm, where the surface conductivity of graphene $\sigma$ can be calculated by the Kubo formula[35-37] (see Methods). The resonant frequency of graphene plasmon (GP) $\omega_{GP}$ is $1553$ cm$^{-1}$ for the doped graphene Fermi energy $E_f = 0.5$ eV (the light green curve in Fig. 1b). We assume that the molecular layer covering the graphene nanoribbon and study the near-field coupling between GP and molecular vibrational modes. The light red curve in Fig. 1c shows that the signals from such a thin molecular layer in the extinction spectra are very weak, even with the enhancement provided by GP. This phenomenon can be understood in terms of coupled harmonic oscillators[39]. Plasmon−phonon coupling generates two new hybrid modes, whose splitting distance and damping depend on their coupling strength and original damping rates $\gamma_{GP}$ and $\gamma_M$. Specially, the damping rates of the hybrid modes are equal to $(\gamma_{GP} + \gamma_M)/2$ when the resonant frequencies of the plasmon and the molecular mode coincide ($\omega_{GP} = \omega_1$). In the case of low concentrations, the hybrid-mode linewidth characterized by $\gamma_{GP}$ and $\gamma_M$ is relatively larger than the splitting distance caused by the weak coupling strength, resulting in a large overlap between the two hybrid-mode broad peaks and a small dip that is difficult to detect in the extinction spectra. Similarly, we use synthesized CFW to recover the molecular signals. Note that even if the decay constant of graphene $\Gamma$ is generally much larger than $\gamma_M/2$, CFW can still partially compensate for $\gamma_{GP}$ (the dark green curve in Fig. 1b), thereby narrowing the linewidths of hybrid modes. It is numerically confirmed that owing to the compensation by synthesized CFW, the originally weak signals are greatly enhanced, and phonon-induced transparency (PIT) structure[40] ($\omega_1$) and Fano structure[41] ($\omega_2$) are clearly visible in the spectra (The dark red curve in Fig. 1c).

We also studied the effect of CFW under different Fermi energies. For the graphene nanoribbon, the resonance of GP gradually shifts to higher frequency with the increase of $E_f$. Due to the relatively large damping, plasmon−phonon coupling has almost no effect on the linear dispersion of GP, such a weak perturbation produces almost no visible dip in the extinction spectrum (Fig. 1d). If we set $\gamma_M$ close to 0, GP dispersion will be strongly affected near the resonance frequencies of the molecular vibrational modes and GP, as shown by the strong anti-crossing behavior (Fig. 1e). We next obtain the spectrum at CFW (Fig. 1f) by applying Eq. 2 to the spectrum at real frequencies (Fig. 1d). The CFW spectrum exhibits strong anti-crossing behavior at the molecular vibrational resonance frequencies, similar to the case of negligible loss (Fig. 1e). Thus, synthesized CFW can effectively enhances GP-based molecular signals through the damping compensation mechanism. In addition, obtaining the phase $\arg(t_M)$ by Kramers–Kronig relations facilitate the applicability of the proposed synthesized CFW method.

**Enhancement of molecular fingerprint signals**

Based on the above theoretical analysis, we take measurements of molecular infrared spectra to showcase the effectiveness of the synthesized CFW method in enhancing sensitivity. In the experiment, Fourier transform infrared (FTIR) spectroscopy is used to measure the molecular infrared vibrational fingerprint spectrum (details in Methods), where the infrared beam excites the molecular vibrations and is absorbed at the specific resonance frequencies (Fig. 2a). We start with deoxynivalenol (DON) molecules, a mycotoxin from Fusarium fungi found in cereals which poses health risks to humans and animals. The optical micrograph (Fig. 2b) illustrates the preparation of DON samples on a Si substrate (details in Methods). It should be noted that the granular shape of DON molecules is mainly due to solvent evaporation and intermolecular interactions.

Due to the large number of molecules in DON particles, the signal intensity after infrared spectroscopy measurement is relatively strong, as shown in the grey curve in Fig. 2c. However, the spin-coated molecules exhibit disorder and have a low-quality factor, resulting in a significant broadening of C-O-H bending modes ($\delta$(C-O-H)) which have fingerprints between 1400-1455 cm$^{-1}$(as indicated by the dashed lines), making them difficult to discern in the extinction spectra. We employed CFW in conjunction with the Kramers–Kronig relation to process the original extinction spectrum (grey curve), obtaining the new spectrum (black curve) in Fig. 2c, clearly displaying the narrowing of the spectral linewidth and enhanced characteristic intensity. This enhancement has allowed us to identify molecular structures and properties more precisely and accurately, contributing significantly to our understanding of molecular spectroscopy.

**Enhancing the sensitivity of graphene-based sensors**

When the molecular layer is thin, or the number of molecules is very small, traditional infrared spectroscopy struggles to effectively probe molecular signals. Currently, graphene-based SEIRA is one of the most sensitive enhanced infrared spectroscopy methods. For implementation, we first soak a graphene-based infrared sensor in a silk protein solution at a concentration of $10\,\mu g/mL$ to enable the protein molecules to adhere to the surface of graphene nanoribbons. The examination of how CFW techniques can enhance the sensors' detection sensitivity is then carried out. Fig. 3a illustrates the schematic of the characterization of GP-enhanced molecular vibrational signal on the periodic graphene nanoribbons. The principle of graphene-based SEIRA is as follows: an infrared light beam irradiates the periodic graphene nanoribbons (Fig. 3b) to excite GP to achieve electromagnetic field enhancement; then, through dynamic back-gate tuning, the resonant frequency of GP is adjusted to be close to the molecular characteristic fingerprint vibrational frequency, resulting in phonon-induced transparency (PIT) in the extinction spectra, as shown in Fig. 3c. The dashed lines indicate the characteristic mode of the protein, representing the Amide I band ($1626\,cm^{-1}$).

We then investigate the extinction spectrum of GP with different thicknesses ($\sim 1.2\,nm$, $\sim 2.1\,nm$, $\sim 3.0\,nm$, and $\sim 5.8\,nm$) of silk proteins on graphene nanoribbons (see details in Figure S3). As the thickness of silk protein increases, there is a corresponding increase in the intensity of the molecular characteristic vibration signal, and the dip of GP-enhanced molecular coupling gradually deepened, as shown in Fig. 3c. However, for silk protein thicknesses of less than 2 nm, the ultra-low coupled signal between graphene and silk protein makes it difficult to identify the vibration signals of the silk protein, which has been a common challenge encountered in the detection of trace proteins. Here, synthesized CFW method is utilized to greatly enhance the signal in the molecular protein growth process, allowing for clear identification of the dips generated at different thicknesses of silk protein, as shown in Fig. 3d. The dip depth $\Delta h$ is used as the figure of merit to quantitatively evaluate the sensitivity of the graphene-based sensor[42,43]. $\Delta h$ is extracted for both real frequency spectra (Fig. 3c) and CFW spectra (Fig. 3d) and plotted in Fig. 3e, demonstrating that the sensitivity has increased by almost one order of magnitude using the CFW method. For the thinnest molecular layer (~1.2 nm), the enhancement factor reaches 15. These results highlight the potential of graphene-based sensors for providing highly sensitive and accurate detection of molecular fingerprints.

**Enhancing the sensitivity of tunable graphene-based liquid phase sensors**

We further apply our method to sensing molecules in aqueous solution, employing a liquid-phase GP-enhanced FTIR experimental setup as depicted in Fig. 4a, which can eliminate the water background outside the GP hotspot. This setup involves a GP-enhanced infrared sensor encapsulated in an infrared-transparent microfluidic system, allowing a transmittance measurement and a steady solution flowing path. In an aqueous environment, the abundant ions form an electric double layer (EDL) on charged surfaces-graphene. Thus, a liquid gate is applied, which enables a stable and tunable plasmon response of the graphene nanoribbons in an aqueous environment. Then, by injecting a bovine serum albumin (BSA) protein solution (1 mg/mL) into the microfluidic system for two hours, protein molecules become saturated and adsorbed onto the graphene nanoribbons. This leads to the appearance of two dips in the extinction spectrum corresponding to the amide I band ($1655~\text{cm}^{-1}$) and amide II band ($1545~\text{cm}^{-1}$) of the BSA protein, as shown in Fig. 4b. The resonant frequency of GP $\omega_{\text{GP}}$ in the infrared fingerprint region can be dynamically adjusted by modulating the doped graphene Fermi energy $E_{\text{f}}$ using the liquid gate $\Delta V_{\text{g}}$. Increasing $\Delta V_{\text{g}}$ from 1.1 V to 2.1 V leads to a blue shift in $\omega_{\text{GP}}$.

Due to significant damping and the presence of noise in an aqueous solution, the two characteristic resonances of the molecule (the dashed curves in Fig. 4b) appear to be indistinct, particularly at larger detuning between GP and the molecular vibrational modes. Here, by applying the CFW method, we observe significantly enhanced signals in the spectrum (the solid curves in Fig. 4b), clearly showing that with an increase of $\Delta V_g$, the detuning first decreases and then increases, causing the line-shapes of the two dips to gradually transform from Fano to PIT, and then back to Fano resonances. At Fano resonances, the dips slightly deviate from the molecular vibrational modes, which is consistent with the theory (Fig. 1c). Moreover, we simulated the extinction in aqueous solution (the map in Fig. 4c). We show that the position of dips in the experimental spectra at CFW (the hollow points in Fig. 4c) conforms to the evolution trend of the simulation, further proving the rationality of CFW method. Therefore, synthesized CFW method is also suitable for enhancing the sensitivity of liquid-phase infrared sensors even under very challenging conditions.

**Conclusions**

In conclusion, we have applied a novel synthesized CFW method to compensate for the intrinsic damping of the detected molecules and sensors, resulting in a large enhancement in the signals of the molecular vibrational fingerprints. We demonstrate that for different experimental scenarios, including DON molecules without plasmonic enhancement and silk protein molecules and BSA protein solutions measured by graphene-based plasmonic sensors, synthesized CFW method can effectively enhance the characteristic signals, exhibiting its wide applications. Importantly, under the condition of low concentrations or thin thicknesses, the CFW method can dramatically improve the signals, which is beneficial to increase the upper limit of sensitivity for various sensors. The CFW technique presents a new platform for the sensing field, enabling the enhancement of sensing sensitivity in complex environments and laying the foundation for environmental monitoring, healthcare diagnosis, and new material development.

## Methods

### Simulation of transmission spectrum

The thin layer system consists of the molecular layer and graphene nanoribbon and is simulated by finite-element method (FEM) using COMSOL Multiphysics software. In the simulation, a transverse magnetic wave is normally incident onto the thin layer with periodic boundary conditions, and then the transmission coefficient is obtained. In addition, the surface conductivity of graphene $\sigma$ used in the simulation is calculated by the Kubo formula,

$$\sigma = \frac{ie^2 E_f}{\pi \hbar^2 (\omega + i\Gamma)} + \frac{ie^2}{4\pi\hbar} \ln \left[ \frac{2|E_f| - (\omega + i\Gamma)}{2|E_f| + (\omega + i\Gamma)} \right]$$

The temperature $T = 300$ K satisfies the approximate requirement of $K_B T \ll E_f$, $e$ is the electron charge, $\omega$ is the angular frequency, $\hbar$ is the reduced Planck constant, and $E_f$ is the doped graphene Fermi energy. The relaxation time $\Gamma = ev_f^2/\mu E_f$, where $v_f = c/300$ is the Fermi velocity and $\mu = 700 \text{ cm}^2\text{V}^{-1}\text{s}^{-1}$ is the carrier mobility of graphene.

### Graphene Plasmonic IR Biosensing

For silk protein detection, the proposed graphene plasmon infrared sensor was composed of connected graphene nanoribbon arrays patterned on a 285 nm SiO$_2$/500 µm substrate (from Silicon Valley Microelectronics, Inc.) using electron beam lithography (Vistec 5000+ES, Germany) and oxygen plasma etching (SENTECH, Germany). The monolayer graphene film was grown on copper foil by chemical vapor deposition and transferred to a SiO$_2$/Si substrate using the wet transfer method. The graphene nanoribbon arrays were designed to have widths ranging from 50 nm to 100 nm, with a gap of 50 nm to 100 nm. A pair of electrodes (5 nm Ti and 50 nm Au) were patterned and evaporated onto the graphene using electron-beam lithography combined with electron beam evaporation (OHMIKER-50B, Taiwan). The back-gate was applied by connecting the electrode to the backside of the SiO$_2$/Si substrate using an external circuit with the help of silver thread. The SourceMeter (Keithley 2636B) was utilized to supply varied gate voltages.

For BSA protein solution detection, based on the previous setup of the graphene plasmon infrared sensor, a top-gate electrode was evaporated onto the substrate, which was connected to the solution but not to the graphene nanoribbons. Additionally, the source and drain electrodes were further passivated with a 50 nm PMMA layer to prevent direct interaction between the electrodes and protein molecules, as well as to minimize electrolyte leakage between the source and gate. The graphene plasmon infrared sensor was then encapsulated with a microfluidic system. Finally, the electrodes were led outside the microfluidic system and connected to the external circuit using silver thread.

**Characterization of the Graphene Plasmon Infrared Sensor**

The morphologies and thicknesses of the fabricated graphene nanoribbons were characterized by employing scanning electron microscopy (NOVA Nano SEM 430) and AFM (Bruker Multimode8) measurements. The transfer characteristic curve was determined by using a source meter (Keithley 2636B). The FTIR transmission measurements were performed with Thermo Fisher Nicolet iN10 with an IR microscope (15× objective). The aperture was set as 100 μm × 200 μm for each measurement, while the resolution was 4 cm$^{-1}$ and scans were 128.

**The chemicals sampling**

The DON solution was prepared by dissolving DON powder in alcohol at a concentration of 0.5 μg/mL. Subsequently, it was dropped onto a Si substrate. Once the alcohol evaporates, the DON molecules remain deposited on the substrate. The 10 μg/mL silk protein solution was prepared by diluting 50 mg/mL silk fibroin solution (from Sigma) 5000 times with deionized water. The BSA protein solution was prepared by dissolving Bovine albumin Fraction V (from Sigma) in deionized water. The different thicknesses of silk protein on graphene nanoribbons were prepared by soaking the graphene plasmon infrared sensor in silk protein solution for varying durations.

## Data availability

The data that support the findings of this study are available from the corresponding author upon reasonable request.

## Acknowledgments

This work was supported by the New Cornerstone Science Foundation, the Research Grants Council of Hong Kong AoE/P-502/20, 17309021 (to S.Z.); National Natural Science Foundation of China (51925203 to Q.D. and 52102160 to X.G.).

## Author contributions

S. Z. and Q.D. conceived the project. K.Z. and X.G. performed numerical simulations. K.Z. and F.G. performed analytical calculations. C.W. conducted the experiments with the help of Y.D. under the supervision of X.Y. and Q.D.; K.Z., C.W., and X.G. processed the experimental data. K.Z., C.W., X.G., F.G., X.Y., N.L., Q.D. and S.Z. analyzed the experimental results. L.L.Z and K.Z. performed the analysis of the signal sidebands under the truncated CFW excitation. K.Z., X.G., C.W., L.L.Z., N.L., Q.D. and S.Z. wrote the manuscript with input from all authors. All authors contributed to the discussion.

## Additional information

Supplementary information is available in the online version of the paper. Reprints and permissions information is available online at www.nature.com/reprints. Correspondence and requests for materials should be addressed to X.Y., N.L., Q.D., S.Z.

## Competing interests

The authors declare no competing financial interests.

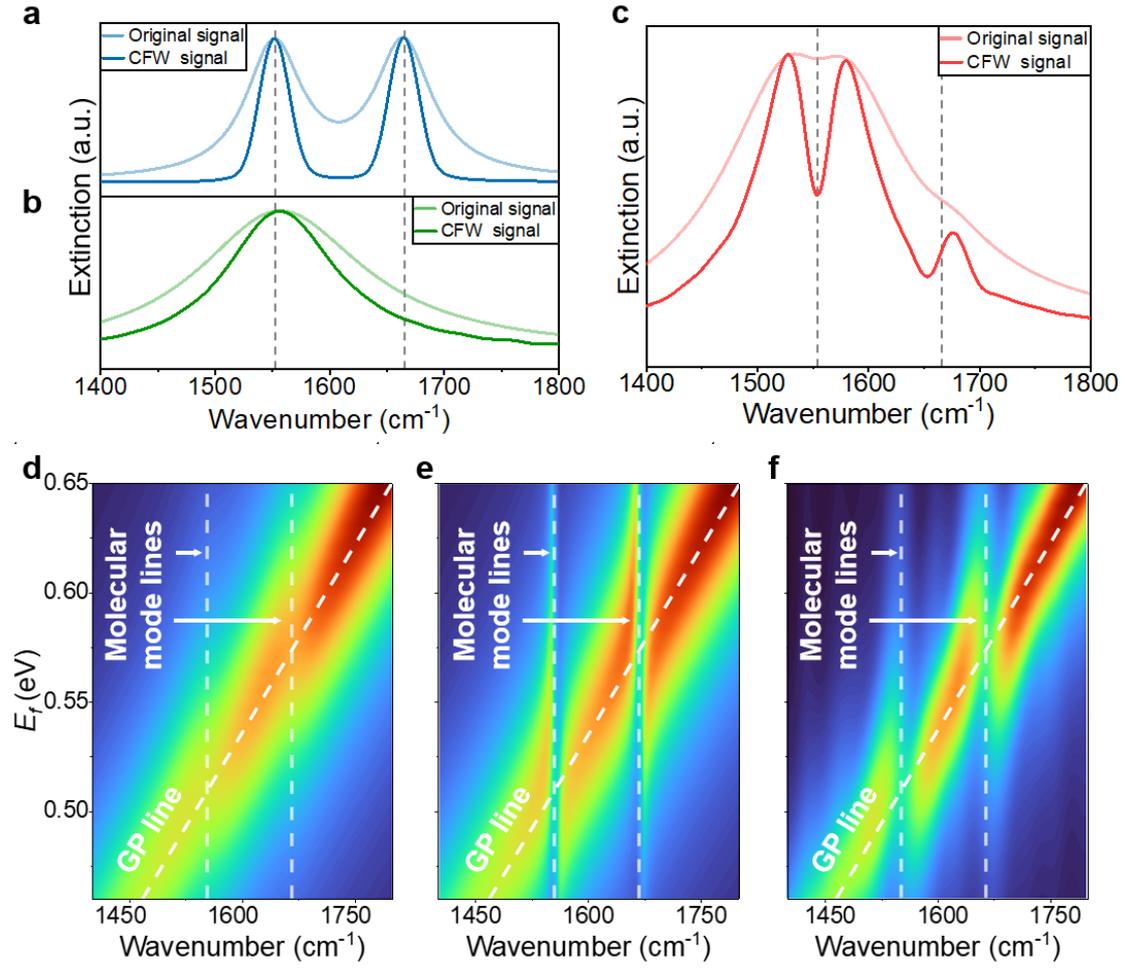

**Fig. 1. Illustration of damping compensation for sensing enhancement through synthesized complex-frequency waves.** (**a**) The extinction spectra of the molecular layer at real frequency (the light blue curve) and at CFW (the dark blue curve). The vibrational peaks are marked by the vertical lines. (**b**) The extinction spectra of the graphene nanoribbon at real frequency (the light green curve) and at CFW (the dark green curve). (**c**) The extinction spectrum of the molecular layer enhanced by GP at real frequency (the light red curve) and at CFW (the dark red curve). (**d-f**) The extinction maps under different $E_f$ (**d**) with $\gamma_M = 60 \ cm^{-1}$ at real frequency, (**e**) with $\gamma_M = 0$ at real frequency and (**f**) with $\gamma_M = 60 \ cm^{-1}$ at CFW. All CFW results are use equation (2) to calculate from the corresponding real frequency results.

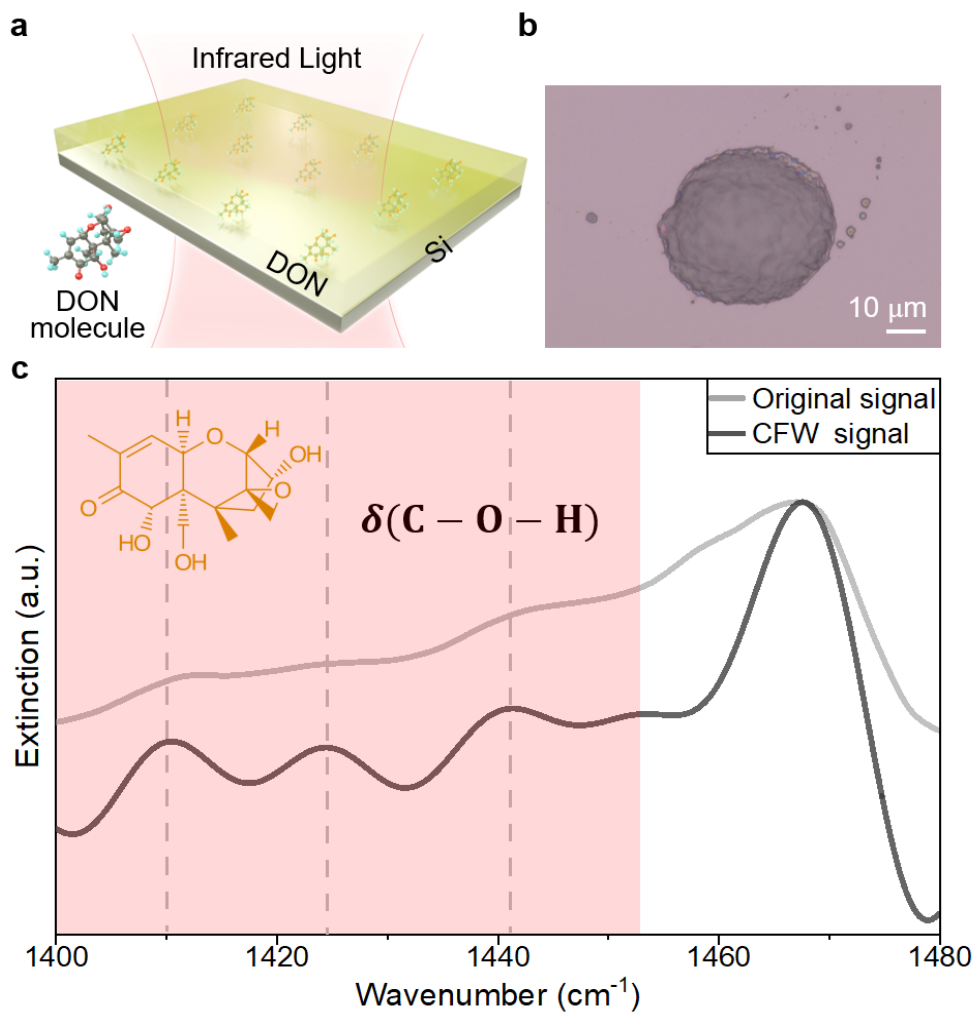

**Fig. 2. Direct observation of molecular vibration infrared signals and the enhancement effect of synthesized complex-frequency waves**. **(a)** Schematic illustration of molecular infrared spectroscopy measurements. The infrared beam is incident upon the molecular film, spin-coated on the Si substrate, to excite the infrared fingerprint vibrational signal. **(b)** Optical microscope image showing the deposition of deoxynivalenol (DON) molecules on Si substrate. After spin coating, the molecules recrystallize into raised particle shapes. Scale bar: 10 um. **(c)** Vibrational extinction spectrum of DON molecules. The gray curve represents the experimentally measured extinction spectrum, while the black curve represents the extinction spectrum enhanced by the complex frequency technique. The vertical lines highlight the characteristic vibrational peaks of the molecules.

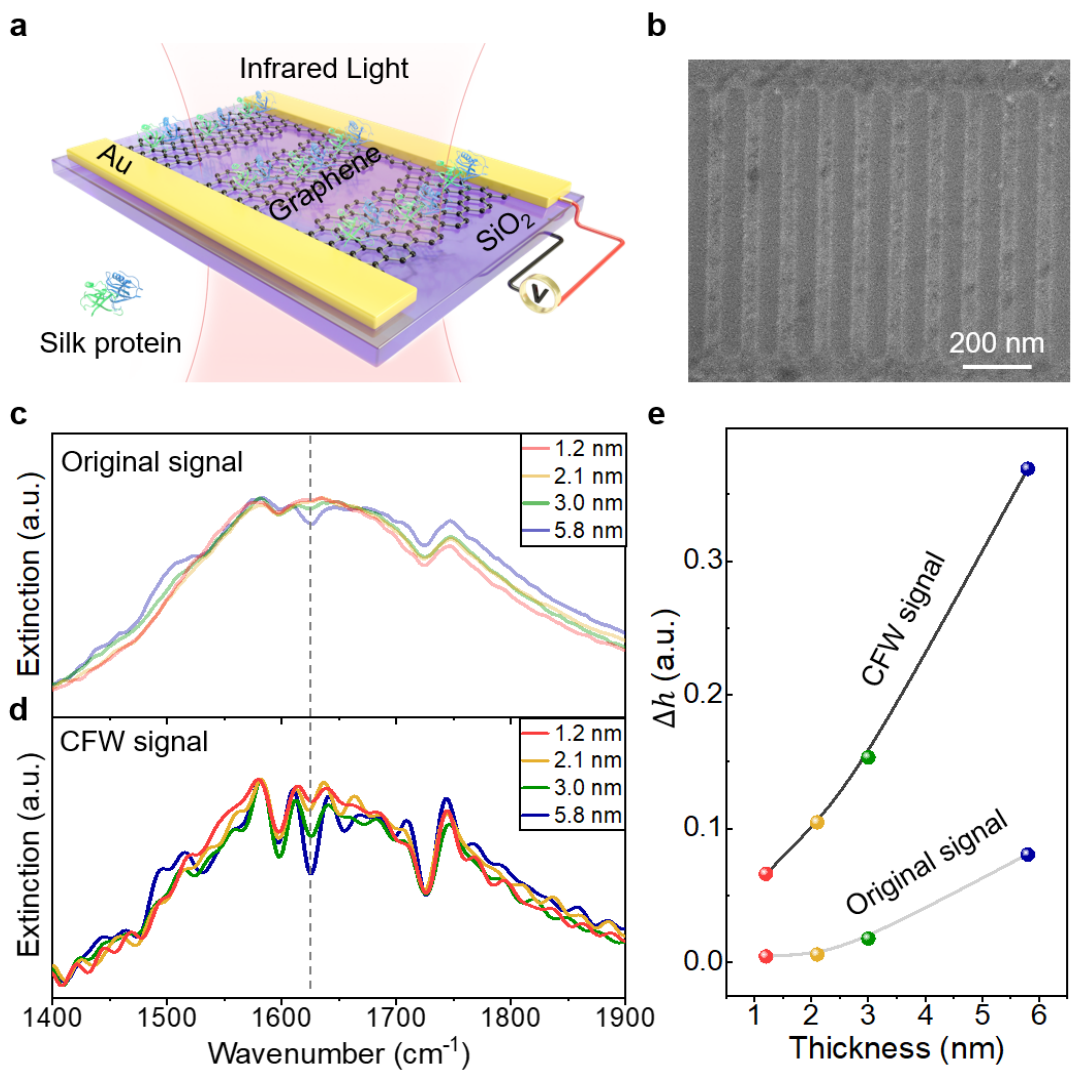

**Fig. 3. Enhancing the sensitivity of graphene-based sensors and monitoring different molecular layer thicknesses**. (**a**) Schematic of the graphene-based sensor. Gate voltage modulation enables GP-enhanced detection of molecular signals. (**b**) The scanning electron microscopy (SEM) image of the graphene-based sensor showcases its distinctive design, characterized by periodic nanoribbons that enhance its sensing capabilities. The width of the graphene nanoribbon is approximately 50 nm, and the period is approximately 100 nm. (**c**) Direct experimental response of the GP infrared sensor with different thicknesses of silk protein (see details in Figure S3) and (**d**) the response enhanced by synthesized CFW. The characteristic vibrational peak of the protein are marked by the vertical line. (**e**) A comparison of the GP-enhanced signals is shown. The sensitivity is scaled by the dip depths $\Delta h$ in (**c-d**).

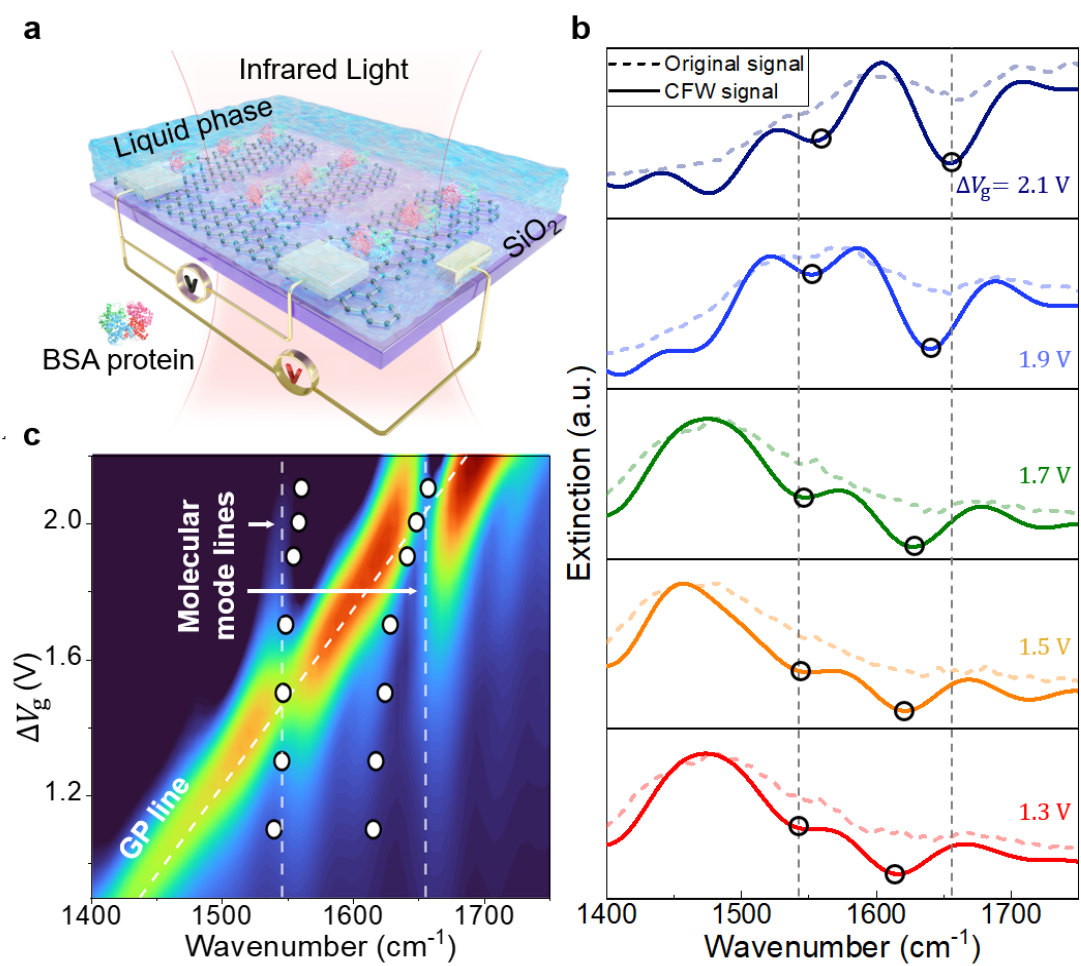

**Fig. 4. Enhancing the sensitivity of liquid phase in situ infrared detection. (a)** Schematic diagram of a liquid-phase GP infrared sensor. The black V is the voltage applied between the source and the drain electrodes, and the red V is the top-gate voltage. **(b)** The dashed curves represent experimental results (see details in Figure S4) of the extinction at different gate voltages $\Delta V_g$, while the solid curves show the signals enhanced by CFW method. The experimental data were collected after 1 mg/mL protein solution flowed for 2 hours, with the graphene nanoribbon width being approximately 50 nm and the period being approximately 120 nm. The characteristic vibrational peaks of the protein are marked by the vertical lines. **(c)** The simulation extinction map under different $\Delta V_g$ by CFW method. The hollow points are the position of dips obtained from experimental results at different $\Delta V_g$ (1.1 V, 1.3 V, 1.5 V, 1.7 V, 1.9 V, 2 V, 2.1 V) by the CFW method.